\documentclass[amsfonts,twocolumn,prl,aps,showpacs]{revtex4}
\usepackage{graphicx}

\begin{document}

\title{Experimental demonstration of a controlled-NOT wave-packet gate}

\author{B. DeMarco}
\author{A. Ben-Kish}
\altaffiliation[Present Address: ]{Dept. of Physics, Technion,
Haifa ISRAEL}
\author{D. Leibfried}
\author{V. Meyer}
\author{M. Rowe}
\altaffiliation[Present Address: ]{NIST Optoelectronics Division}
\author{B.M. Jelenkovi\'{c}}
%\altaffiliation[Present Address: ]{Institute of Physics, Belgrade,
%Yugoslavia}
\author{W.M. Itano}
\author{J. Britton}
\author{C. Langer}
\author{T. Rosenband}
\author{D.J. Wineland}
\address{NIST Boulder, Time and Frequency Division, Ion Storage Group}
\date{\today}

\begin{abstract}

We report the experimental demonstration of a controlled-NOT
(CNOT) quantum logic gate between motional and internal state
qubits of a single ion where, as opposed to previously
demonstrated gates, the conditional dynamics depends on the extent
of the ion's wave-packet. Advantages of this CNOT gate over one
demonstrated previously are its immunity from Stark shifts due to
off-resonant couplings and the fact that an auxiliary internal
level is not required.  We characterize the gate logic through
measurements of the post-gate ion state populations for both logic
basis and superposition input states, and we demonstrate the gate
coherence via an interferometric measurement.
\end{abstract}

\maketitle

Considerable attention is now focused on developing quantum
computer technology across a diverse set of physical systems
\cite{QC1,QC2}. Using laser cooled trapped atomic ions for quantum
computing and information processing carries many advantages as
most of the basic building blocks of quantum computing have been
demonstrated, including high efficiency state initialization and
read-out \cite{Monroe95a,Meekhof96,King99,Ben-Kish02}, entangling
gates \cite{Monroe95b,Sackett00}, individual addressing
\cite{Nagerl99}, and long qubit coherence times
\cite{Wineland98,Roos99}. Areas of concentration for current work
with trapped ions include simplifying quantum logic operations and
increasing their fidelity, as well as scaling up the complexity of
computations \cite{Sackett01,Rowe02,Kielpinski02}. Here we report
the experimental demonstration of a CNOT logic gate between
motional and internal-state qubits of a trapped $^9$Be$^+$ ion
that requires fewer resources than that of Ref. \cite{Monroe95b}.
This gate is fundamentally different than previously demonstrated
gates in that the conditional dynamics depend on the size of the
atomic wave-packet compared to the wavelength of the applied
radiation \cite{Monroegate}.

The quantum CNOT logic gate between two qubits has become a
paradigm for quantum computing because universal quantum
computation can be carried out with the CNOT gate and single qubit
rotations \cite{QC1}. A CNOT gate toggles the state of a target
qubit depending on the state of a control qubit. If the logic
basis states are labelled as $\left|0\right\rangle$ and
$\left|1\right\rangle$, then the general target state
\mbox{$\cos(\theta)\left|0\right\rangle+e^{i\phi}\sin(\theta)\left|1\right\rangle$}
should be unaffected if the control qubit is in the
$\left|0\right\rangle$ state and transformed to
\mbox{$\cos(\theta)\left|1\right\rangle+e^{i\phi}\sin(\theta)\left|0\right\rangle$}
for a $\left|1\right\rangle$ control qubit.  Here $\theta$ and
$\phi$ are taken to be arbitrary angles to designate the most
general superposition state. Based on a previous proposal
\cite{Monroegate}, we demonstrate a CNOT gate in the trapped ion
system that requires only a single laser pulse and no auxiliary
levels, and is therefore simplified compared to a previous
implementation that required three laser pulses and an auxiliary
internal state \cite{Monroe95b}. Furthermore, the method employed
here is free from level shifts introduced by the gate coupling.

The qubits in our implementation of the CNOT gate are spanned by
the internal and motional levels of a single trapped $^9$Be$^+$
ion \cite{Monroe95b,Sackett01}. The target states, abbreviated by
the analogous spin-$1/2$ states $\left|\downarrow\right\rangle$
and $\left|\uparrow\right\rangle$, are the
\mbox{$\left|F\!=\!2,m_F\!=\!-2\right\rangle$} and
\mbox{$\left|F\!=\!1,m_F\!=\!-1\right\rangle$} hyperfine states of
the $^9$Be$^+$ $^2S_{1/2}$ electronic ground state.  The control
qubit consists of the ground and second excited states
($\left|0\right\rangle$ and $\left|2\right\rangle$) of the
quantized states $\left|n\right\rangle$ of the harmonic ion motion
along the trap axis or $z$ direction.  Coherent manipulation of
the four qubit states and the CNOT gate action is accomplished by
driving two-photon Raman transitions using two laser beams detuned
from the \mbox{$^2S_{1/2}\rightarrow^2P_{1/2},^2P_{3/2}$}
transitions \cite{Monroe95b,Sackett01}. The laser beams can be
used to change just the spin state of the ion by driving the
``carrier" transition
\mbox{$\left|\downarrow\right\rangle\left|n\right\rangle\leftrightarrow\left|\uparrow\right\rangle\left|n\right\rangle$}
with a laser beam frequency difference
\mbox{$\omega_0\sim2\pi\times1.25$}~GHz. By tuning the frequency
difference to $\omega_0\pm\Delta n\cdot\omega_z$ (where $\omega_z$
is the harmonic oscillator frequency) both the spin state and
motional level can be changed via driving the $\Delta n^{th}$
order sideband:
\mbox{$\left|\downarrow\right\rangle\left|n\right\rangle\leftrightarrow\left|\uparrow\right\rangle\left|n+\Delta
n\right\rangle$}. We refer to the ``blue" sidebands for
\mbox{$\Delta n\ge1$} and ``red" sidebands for \mbox{$\Delta
n\le-1$}. In this way, any of the four qubit eigenstates
($\left|\downarrow\right\rangle\left|0\right\rangle$,
$\left|\uparrow\right\rangle\left|0\right\rangle$,
$\left|\downarrow\right\rangle\left|2\right\rangle$,
$\left|\uparrow\right\rangle\left|2\right\rangle$) and any
superposition of them can be generated from the initial state
$\left|\downarrow\right\rangle\left|0\right\rangle$ that is
prepared by optical pumping and laser cooling \cite{Meekhof96}.

The CNOT gate is implemented by applying a single Raman laser
pulse on the carrier transition.  The gate action relies on tuning
the laser-atom interaction by adjusting the relative overlap of
the motional qubit states with the laser field. The carrier Rabi
rate for the $\left|2\right\rangle$ motional state is reduced
compared to the $\left|0\right\rangle$ state because the
$\left|2\right\rangle$ state has a broader spatial extent and the
ion averages over the laser wave \cite{Wineland98,debyenote}. This
effect is a manifestation of the wave-packet nature of the ions --
we cannot obtain the observed interaction with the laser by
assuming that the ions are point particles \cite{suppression}.

\begin{figure}[h!]
\includegraphics[scale=0.5]{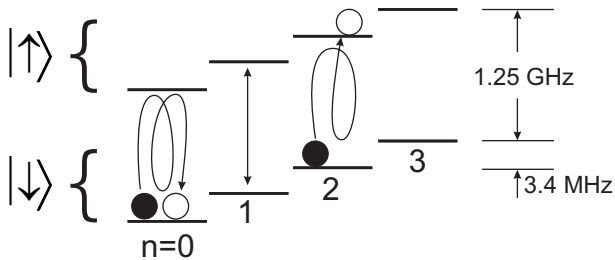}
\caption{\label{fig1}Schematic of the trapped ion levels and CNOT
gate operation.  Shown here are the trapped ion motional levels
(separated in energy by $h\times3.4$~MHz, where $h$ is Planck's
constant) and internal spin states $\left|\uparrow\right\rangle$
and $\left|\downarrow\right\rangle$.  The motional states act as
the control and the internal state as the target qubit for the
gate. The control qubit is composed of the ground and second
excited motional states ($n=0,2$) along the trap axis. The CNOT
gate is operated by driving the carrier transition which couples
spin states with the same $n$. The laser-atom interaction is tuned
so that an ion in the $\left|0\right\rangle$ state returns to its
initial spin state while an ion in the $\left|2\right\rangle$
state toggles its spin state when the gate is driven. In the
figure, filled circles represent two possible input ion
eigenstates and open circles show the corresponding output state
after the CNOT gate is applied.}
\end{figure}

By adjusting the trap strength and therefore manipulating the
Lamb-Dicke parameter $\eta$, the ratio of the carrier transition
Rabi rates for the $\left|0\right\rangle$ and
$\left|2\right\rangle$ states
\mbox{$\Omega_{0,0}/\Omega_{2,2}=2/(2-4\eta^2+\eta^4)$} is set to
4/3 in the experiments reported here. Here $\Omega_{i,j}$ is the
two-photon Rabi rate for the coupling
\mbox{$\left|\downarrow\right\rangle\left|i\right\rangle\leftrightarrow\left|\uparrow\right\rangle\left|j\right\rangle$}
and $\eta$ depends on the trap frequency through $\eta=\Delta
k_z\sqrt{\hbar/2m\omega_z}$ where $m$ is the ion mass and $\Delta
k_z$ is the wavevector difference of the Raman laser beams along
the $z$ direction \cite{Wineland98}. To operate the gate, the
carrier transition is driven for a time $t_{\sf gate}$, evolving
the state amplitudes $c$ according to \mbox{$c'_{\uparrow
n}=\cos(\Omega_{n,n} t_{\sf gate})c_{\uparrow
n}-i\sin(\Omega_{n,n} t_{\sf gate})c_{\downarrow n}$} and
\mbox{$c'_{\downarrow n}=\cos(\Omega_{n,n} t_{\sf
gate})c_{\downarrow n}-i\sin(\Omega_{n,n} t_{\sf gate})c_{\uparrow
n}$} (cf. Eq. 23 of \cite{Wineland98}). The amplitudes $c$ are
defined so that the state of the ion is given by $\sum_{n=0,2}
\left(c_{\downarrow n}\left|\downarrow\right\rangle+c_{\uparrow
n}\left|\uparrow\right\rangle\right)\left|n\right\rangle$. The
pulse time $t_{\sf gate}$ is chosen so that the
$\left|0\right\rangle$ state undergoes two full Rabi cycles, or a
``$4\pi$-pulse" ($\Omega_{0,0} t_{\sf gate}=2\pi$). For the same
pulse time the $\left|2\right\rangle$ state experiences 1.5 Rabi
cycles, or a ``$3\pi$-pulse" ($\Omega_{2,2} t_{\sf gate}=1.5\pi$).
Under these conditions, the target bit (spin) of the atom flips if
the control bit (motional state) is in the $\left|2\right\rangle$
state and stays the same for the $\left|0\right\rangle$ state,
accomplishing the CNOT gate logic (see Fig. \ref{fig1}). The
$\pi/2$ phase acquired on the $\left|2\right\rangle$ state can be
removed by an appropriate phase shift in a subsequent operation.
The scheme employed here is a specific case of more general
possibilities outlined in \cite{Monroegate}.

A single $^9$Be$^+$ ion is trapped in a linear Paul trap using a
combination of static and time varying electric fields to produce
a 3-dimensional harmonic confining potential \cite{Rowe02}.  The
harmonic oscillator frequency along the trap ($z$) axis is set to
3.4~MHz by adjusting the static potentials in order to adjust
$\eta$ so that \mbox{$\Omega_{0,0}/\Omega_{2,2}=4/3$}. State
preparation and qubit manipulation are accomplished using a pair
of non-collinear Raman beams \cite{Meekhof96,Sackett01}.  The
beams are detuned by $\sim+80$~GHz from the electronic $^2S_{1/2}$
to $^2P_{1/2}$ transition at 313~nm, and the intensity in the
beams is set to give $\Omega_{0,0}=2\pi\times92$~kHz.

Each experiment begins by initializing the ion qubit into the
$\left|\downarrow\right\rangle\left|0\right\rangle$ state with
99.9\% probability using resolved-sideband Raman cooling and
optical pumping \cite{Monroe95a,Leibfried97,King99}. The gate
input state is then prepared using combinations of carrier and
sideband transitions. After applying the gate, the experimental
observable is the spin state of the ion which is determined
through resonance fluorescence measurements
\cite{Rowe01,Sackett01}. The ion is illuminated for 200~$\mu$s by
a $\sigma^{-}$ polarized beam detuned by -10~MHz from the
electronic $^2S_{1/2}$, $\left|\downarrow\right\rangle$ to
$^2P_{3/2}$ ($m_J=-3/2$) excited state transition.  The histogram
of scattered photons collected onto a photo-multiplier tube (PMT)
is recorded for 200 experiments with identical parameters. The
bright $\left|\downarrow\right\rangle$ state is distinguished from
the dark $\left|\uparrow\right\rangle$ state by the difference in
photon scattering rates. The probability
$P_\downarrow=\sum_n\left|c_{\downarrow n}\right|^2$ to find the
ion in the $\left|\downarrow\right\rangle$ state is determined by
fitting the measured fluorescence histogram to reference
histograms measured from the
$\left|\downarrow\right\rangle\left|0\right\rangle$ and
$\left|\uparrow\right\rangle\left|0\right\rangle$ states.  The fit
uncertainty is typically less than 1\%.

We characterize the logic gate by measuring both the logic truth
table and the gate coherence.  The CNOT gate truth table is
determined by measuring the ion spin output state for different
input eigenstates, in analogy with a classical logic gate. The
measured output state for each of the logic basis states is shown
in table \ref{truthtable}. The four basis states are generated
from the $\left|\downarrow\right\rangle\left|0\right\rangle$ state
using combinations of carrier and sideband $\pi$-pulses
\cite{Meekhof96}: the
$\left|\uparrow\right\rangle\left|0\right\rangle$ state is
prepared from the
$\left|\downarrow\right\rangle\left|0\right\rangle$ by driving a
$\pi$-pulse on the carrier transition, the
$\left|\downarrow\right\rangle\left|2\right\rangle$ state is
prepared by driving a $\pi$-pulse on the first blue sideband
(\mbox{$\Delta n=+1$}) and then first red sideband (\mbox{$\Delta
n=-1$}), and the $\left|\uparrow\right\rangle\left|2\right\rangle$
state is prepared by driving a $\pi$-pulse on the second blue
sideband (\mbox{$\Delta n=+2$}). The input spin state of the ion
is prepared with better than 96\% accuracy. After input state
preparation, the CNOT gate is applied by driving the carrier
transition for $t_{\sf gate}$, and then the ion spin state is
measured. For each of the basis states, we achieve at least 95\%
accuracy in the CNOT logic \cite{off_resonant}.

\begin{table}[h!]
\begin{tabular}{c||c|c|}
 & \parbox[c][20pt][c]{30pt}{\large $\downarrow$} & \large $\uparrow$ \\ \hline\hline
\parbox[c][20pt][c]{30pt}{\large n=0} & ~$0.989\pm0.006$~ & ~$0.050\pm0.007$~ \\
\hline
 \parbox[c][20pt][c]{30pt}{\large n=2} & ~$0.019\pm0.007$~ & ~$0.968\pm0.007$~ \\ \hline
\end{tabular}
\caption{Measured CNOT logic truth table.  The measured
probability that the ion is in the $\left|\downarrow\right\rangle$
state after application of the CNOT gate is shown for different
input eigenstates.  We observe the expected CNOT behavior where
the spin state of the ion is flipped for n=2 and remains unchanged
for n=0 (note that the probabilities do not sum to unity because
these data represent the results of four separate experiments).
 The errors in the gate logic compared to the ideal case include errors in input state initialization.}\label{truthtable}
\end{table}

\begin{figure}
\includegraphics[scale=0.85]{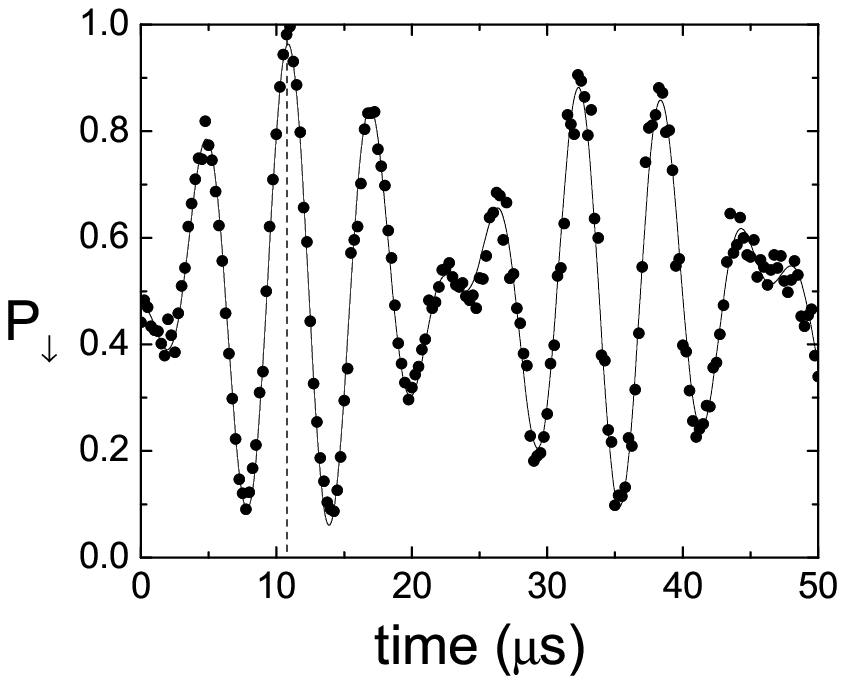}
\caption{\label{fig2}CNOT gate acting on a superposition state.
The input state
\mbox{$\frac{1}{\sqrt{2}}\left(\left|\downarrow\right\rangle\left|0\right\rangle+\left|\uparrow\right\rangle\left|2\right\rangle\right)$}
is prepared and then the carrier transition is driven for an
increasing period of time. At the CNOT gate time $t_{\sf gate}$
(indicated by the dashed line), the probability $P_\downarrow$ to
find the ion in the $\left|\downarrow\right\rangle$ state is 98\%,
which is in agreement with the expected output state
\mbox{$\frac{1}{\sqrt{2}}\left(\left|\downarrow\right\rangle\left|0\right\rangle+i\left|\downarrow\right\rangle\left|2\right\rangle\right)$}.
The solid line is a fit of the data to the sum of two sine
functions with an exponentially decaying envelope. The decay in
the Rabi oscillation contrast with a $170\pm10$~$\mu$s time
constant (as determined by the fit) is primarily due to laser
intensity and magnetic field fluctuations.}
\end{figure}

A key feature of quantum logic gates is the ability to ``parallel
process" superposition input states, a capability that classical
logic lacks. Figure 2 shows the gate acting on the superposition
input state
\mbox{$\frac{1}{\sqrt{2}}\left(\left|\downarrow\right\rangle\left|0\right\rangle+\left|\uparrow\right\rangle\left|2\right\rangle\right)$}.
The input state is prepared by applying a $\pi/2$-pulse on the
second blue sideband starting from the
$\left|\downarrow\right\rangle\left|0\right\rangle$ state. The
CNOT gate should flip the spin if the ion is in the
$\left|2\right\rangle$ state, so that the expected output state is
\mbox{$\frac{1}{\sqrt{2}}\left(\left|\downarrow\right\rangle\left|0\right\rangle+i\left|\downarrow\right\rangle\left|2\right\rangle\right)$}
and the ion is always found in the $\left|\downarrow\right\rangle$
state. Figure \ref{fig2} shows the measured probability to find
the ion in the $\left|\downarrow\right\rangle$ state as the
carrier transition is driven for an increasing period of time. The
beating observed is due to the two different carrier Rabi
oscillation frequencies for the $\left|0\right\rangle$ and
$\left|2\right\rangle$ state. A fit of the data in figure 2 to a
sum of two sine functions gives
$\Omega_{0,0}/\Omega_{2,2}=1.295\pm0.002$. Our ability to set the
desired ratio $\Omega_{0,0}/\Omega_{2,2}=4/3$ is limited in part
by slow drift in $\omega_z$ caused by changing stray electric
fields with spatial curvature.  The accuracy of the gate logic is
robust to deviations from $\Omega_{0,0}/\Omega_{2,2}=4/3$, and, at
the gate time, the ion is in the $\left|\downarrow\right\rangle$
state 98\% of the time.

\begin{figure}
\includegraphics[scale=0.9]{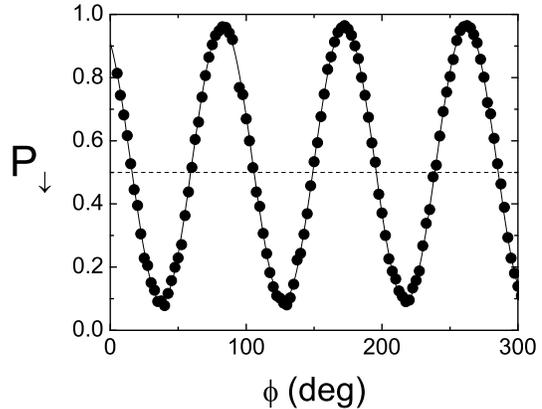}
\caption{\label{fig3}Coherence measurement.  To establish the CNOT
gate coherence, we perform an interferometric phase measurement
using the input state
$\frac{1}{\sqrt{2}}\left|\downarrow\right\rangle\left(\left|0\right\rangle+e^{i\phi}\left|2\right\rangle\right)$.
After applying the CNOT gate and then an analysis $\pi/2$-pulse on
the second blue sideband, the probability to find the ion in the
$\left|\downarrow\right\rangle$ state is measured for different
phases $\phi$.  No sensitivity to $\phi$ is expected if the gate
acts incoherently (shown by the dashed line).  The solid line
shows a fit of the data to a sine function.}
\end{figure}

Measuring the population in the $\left|\downarrow\right\rangle$
state after applying the gate to a superposition input state does
not determine that the CNOT gate acts coherently.  To verify the
gate coherence, we perform an interferometric phase measurement
using the input state
\mbox{$\frac{1}{\sqrt{2}}\left|\downarrow\right\rangle\left(\left|0\right\rangle+e^{i\phi}\left|2\right\rangle\right)$}.
This state is prepared from
$\left|\downarrow\right\rangle\left|0\right\rangle$ by applying a
$\pi/2$-pulse on the first blue sideband with a phase $\phi$
followed by a $\pi$-pulse on the first red sideband. If the gate
acts coherently on the input state, then the pure state
\mbox{$\frac{1}{\sqrt{2}}\left(\left|\downarrow\right\rangle\left|0\right\rangle+ie^{i\phi}\left|\uparrow\right\rangle\left|2\right\rangle\right)$}
is generated by the gate.  Alternatively, if the gate transfers
population incoherently then the state after the gate is
characterized by the mixed state density matrix
$\rho=\frac{1}{2}\left(\left|\downarrow\right\rangle\left|0\right\rangle\left\langle0\right|\left\langle\downarrow\right|+\left|\uparrow\right\rangle\left|2\right\rangle\left\langle2\right|\left\langle\uparrow\right|\right)$.
To test for the coherence, after applying the CNOT gate to the
input state a $\pi/2$ analysis pulse on the second blue sideband
(with arbitrary but constant phase) is applied and then
$P_\downarrow$ is measured. For coherent gate action, the final
state after the analysis pulse should oscillate fully between the
states $\left|\downarrow\right\rangle\left|0\right\rangle$ and
$\left|\uparrow\right\rangle\left|2\right\rangle$, while
incoherent gate behavior produces no dependence on $\phi$. Figure
\ref{fig3} shows the observed oscillations in $P_{\downarrow}$ as
the phase $\phi$ is varied demonstrating the coherence of the
gate. The lack of 100\% contrast in the fringes is due to
imperfections in the state and analysis pulses as well as limited
gate fidelity.

A novel feature of the gate demonstrated here is the absence of
level shifts introduced by the gate coupling.  Other CNOT and
phase gates for the trapped ion system are affected by shifts in
the energy of the levels involved in the gate. These level shifts
can be caused by the existence of off-resonant couplings to
``spectator" levels, for example \cite{Wineland98,Steane00}. If
the energy spacing between levels changes when the coupling is
applied, a phase error can accumulate over the course of an
extended computation which then must be corrected.  In the limit
where the two-photon Rabi rate is small compared to the trap
frequency, the energy shift $\Delta E$ for
$\left|\downarrow\right\rangle\left|n\right\rangle$ and
$\left|\uparrow\right\rangle\left|n\right\rangle$ caused by the
sideband couplings can be expressed as:
\begin{equation}
\Delta E(\downarrow,n)=\hbar\cdot\sum_{i\neq n,i\geq0}
\frac{1}{\omega_z}\frac{\Omega_{i,n}^2}{n-i}=\Delta
E(\uparrow,n)\label{shifteq}
\end{equation}
Here, the shift $\Delta E$ is caused by the presence of
off-resonant sideband couplings to higher and lower energy
motional states. Pairs of coupled levels
($\left|\downarrow\right\rangle\left|n\right\rangle$ and
$\left|\uparrow\right\rangle\left|n\right\rangle$) shift in energy
by the same amount, so that the difference $\Delta
E(\uparrow,n)-\Delta E(\downarrow,n)$ relevant to quantum logic
operations vanishes. Physically this occurs because there are an
equal number of equally detuned red and blue sideband transitions
from states with the same $n$ \cite{error}.  Equation
\ref{shifteq} is valid only when coupling to non-resonant
spectator levels can be considered as a perturbation.  As
$\Omega_{n,n}$ increases, the amplitudes of the spectator levels
become significant so that after applying the gate information is
lost to states outside the computational basis
\cite{off_resonant}. Therefore, as with the gate of Refs.
\cite{Monroe95b} and \cite{CZ}, the gate speed
($\propto\Omega_{n,n}$) must be kept below the ion oscillation
frequency.

In conclusion, using a single trapped $^9$Be$^+$ ion we have
demonstrated a CNOT quantum logic gate between a motional and a
spin qubit that is simplified compared to previous
implementations.  To perform a CNOT gate between two ions, the
state of the control ion must first be mapped onto the selected
motional mode, followed by a CNOT operation between the motion and
target ion as in the original proposal of Cirac and Zoller
\cite{CZ}. However, the gate shown in this work does not require
auxiliary levels, uses only a single laser pulse, and is free from
level shifts caused by the gate coupling. The ability to
manipulate the motional states of the ion was enabled in part by
improvements in ion trap technology;  the latest generation of
NIST ion traps exhibits a factor of 100 reduction in the motional
state heating rate \cite{Rowe02}. With anticipated further
reductions in the heating rate and improvements in system
stability, the conditions for fault-tolerant computation with
trapped ions appears feasible.

The authors thank Murray Barrett and David Lucas for suggestions
and comments on the manuscript.  This work was supported by the
National Security Agency (NSA) and Advanced Research and
Development Activity (ARDA) under contract No. MOD-717.00. This
paper is a contribution of the National Institute of Standards and
Technology and is not subject to U.S. copyright.


\begin{references}

\bibitem{QC1} M.A. Nielsen and I.L. Chuang, {\it Quantum Computation and
Quantum Information}, Cambridge, 2000.

\bibitem{QC2} D. Bouwmeester, A. Ekert, A. Zeilinger, {\it The Physics
of Quantum Computation}, Springer-Verlag, 2001.

\bibitem{Monroe95a} C. Monroe {\it et al.}, Phys. Rev. Lett. {\bf
75}, 4011 (1995).

\bibitem{Meekhof96} D.M. Meekhof, C. Monroe, B.E. King, W.M. Itano, and D.J. Wineland, Phys. Rev. Lett. 76, 1796 (1996).

\bibitem{King99} B.E. King {\it et al.}, Phys. Rev. Lett. {\bf
83}, 4713 (1999).

\bibitem{Ben-Kish02} A. Ben-Kish {\it et al.}, submitted for
review.

\bibitem{Monroe95b} C. Monroe {\it et al.}, Phys. Rev. Lett. {\bf
75}, 4714 (1995).

\bibitem{Sackett00} C.A. Sackett {\it et al.}, Nature {\bf 404},
256 (2000).

\bibitem{Nagerl99} H. C. N\"{a}gerl {\it et al.}, Phys. Rev. A {\bf60}, 145
(1999).

\bibitem{Wineland98} D. Wineland {\it et al.}, J. Res. Natl. Inst. Stand. Technol.
{\bf 103}, 259 (1998).

\bibitem{Roos99} Ch. Roos {\it et al.}, Phys. Rev. Lett. {\bf83}, 4713 (1999).

\bibitem{Sackett01} C.A. Sackett, Quant. Inf. Comp. {\bf 1}, 57 (2001).

\bibitem{Rowe02} M.A. Rowe {\it et al.}, Quant. Inf. Comp. {\bf 4}, 257
(2002).

\bibitem{Kielpinski02} D. Kielpinski, C. Monroe, D.J. Wineland,
Nature {\bf 417}, 709 (2002).

\bibitem{Monroegate} C. Monroe {\it et al.}, Phys. Rev. A {\bf
55}, R2489 (1997).

\bibitem{debyenote} D. J. Wineland and W. M. Itano, Phys. Rev. A {\bf 20}, 1521 (1979).

\bibitem{suppression} A suppression of the Rabi rate due to
radio-frequency micromotion was observed in Q.A. Turchette {\it et
al.}, Phys. Rev. Lett {\bf 81}, 3631 (1998); however, this
suppression was not state dependent and would apply to a point
particle.

\bibitem{Leibfried97} D. Leibfried {\it et al.},  J. Mod. Optics 44, 2485 (1997).

\bibitem{Rowe01} M.A. Rowe {\it et al.}, Nature {\bf 409} 791 (2001).

\bibitem{off_resonant} We estimate for the experimental conditions described in
this manuscript that less than 0.05\% of the ion population is
moved outside of the $n$=0,2 levels by off-resonant excitation of
sideband transitions. This effect can be minimized through an
appropriate choice for the intensity envelope of the laser pulse.

\bibitem{Steane00} A. Steane {\it et al.}, Phys. Rev. A {\bf 62},
042305 (2000).

\bibitem{error} The level shifts introduced by the CNOT gate demonstrated in this work were analyzed in
\cite{Steane00}, but the authors did not consider the balancing
contributions that cause $\Delta E(\uparrow,n)-\Delta
E(\downarrow,n)$ to vanish.

\bibitem{CZ} J. I. Cirac and P. Zoller, Phys. Rev. Lett. {\bf 74}, 4091 (1995).

\end{references}
\end{document}